\documentclass{iau_FM}
\usepackage{graphicx}

\title[Multiwavelength intraday variability in blazars] 
{Multiwavelength intraday variability:  \\what do the studies tell us about\\ blazar jets?} 

\author[Gopal Bhatta ]   
{Gopal Bhatta$^1$}

\affiliation{$^1$Astronomical Observatory of the Jagiellonian University, \\ ul. Orla 171, 30-244 Krak\'ow, Poland \\ email: {\tt gopal@oa.uj.pl}}

\pubyear{2018}
\jname{Astronomy in Focus, Volume 1} 
\editors{Volker Beckmann \& Claudio Ricci, eds.}
\begin{document}
\maketitle

\begin{abstract}
 In this presentation, we report the results of intraday variability in the optical (BVRI bands) and hard X-ray band (3-79 keV) in a number of blazars. In the optical microvariability studies of the blazars S5 0716+714 and BL Lac, we observed many interesting features such as rapid variability, large variability amplitude, presence of characteristic timescales, bluer-when-brighter achromatic behavior, and single power-law power spectral density.  In {\it NuSTAR} observations of several blazars, using spectral and timing analysis, we found similar features consistent with the optical studies. In addition, in BL Lacs we estimated the Lorentz factor of the population of highest energy electrons emitting synchrotron emission, and whereas in flat-spectrum radio quasars, using external Compton models, we estimated the energy of the lower end of the injected electrons to be a few tens of Lorentz factors. In addition, we find that the low flux state exhibit more rapid variability in contrast to the previously reported results showing high flux states displaying rapid variability. In both the studies, the size of the emission regions estimated using variability timescales turn out to be an order magnitude smaller than the gravitational radius of a typical black-hole masses between $\sim 10^8-10^9$ solar masses, believed to be at the center of the radio-loud AGN. The results of the studies suggest that these low-amplitude rapid variability might originate as a result of magnetohydrodynamical instabilities near the base of the jets triggered by the processes modulated by the magnetic field at accretion disc.
\end{abstract}

\firstsection 
\section{Introduction}
Blazars are the radio-loud active galactic nuclei (AGN) that have relativistic jets,  beamed upon us, producing non-thermal emission covering a wide electromagnetic spectrum - from radio to most energetic $\gamma$-rays.  The broadband emission possesses a double-peaked feature in the frequency-flux plane. The lower peak, lying between the radio and the X-ray, is associated with the synchrotron emission by the energetic particles accelerating in the jet magnetic field; whereas the high frequency peak, mostly lying between UV to $\gamma$-ray, is believed due to inverse-Compton scattering of low-frequency photons by high-energy particles. Blazars consist of flat-spectrum radio quasars (FSRQ) and  BL Lacertae (BL Lac) objects. FSQRs, the more powerful sources that show emission lines over the continuum,  have the synchrotron peaks in the lower part of the spectrum; and BL Lacs, the less powerful ones which show weak or no emission lines, have their synchrotron peaks in the higher part of the spectrum. Blazars display variability on diverse timescales ranging from a few minutes to decades. In particular, low amplitude intraday variability, most likely originating at the innermost blazar regions, are central to understanding of blazar physical processes in the vicinity of supermassive black holes in AGN. 

\section{Intraday variability studies}

Over past several years, we closely studied several blazar including S5 716+715, BL Lac and Mrk 501.
In particular, we studied S5 716+715 in two WEBT campaigns using photo polarimetric observations (see \cite[Bhatta et al.(2013)]{bhatta13}, \cite[Bhatta et al.(2015)]{Bhatta2015} and \cite[Bhatta et al.(2016b)]{bhatta16}); and similarly we investigated blazar BL Lac with multi-band observations for several nights (\cite[Bhatta \& Webb(2018)]{Bhatta2018a}).  In addition, several blazars from the NuSTAR data archive were studied (see \cite[Bhatta et al.(2018)]{Bhatta2018b}). The studied were conducted with an aim to characterize the flux and spectral variability in blazar in hour-like timescales. In particular, we searched for characteristic timescales, including minimum variability timescales, employing a number of time series methods e.g., power spectral density (PSD), structure function and auto-correlation function. Furthermore, the spectral analysis were carried out exploring  intra-day spectral evolution and flux spectra correlation.  In hard X-ray observations we studied spectral fitting and evolution of hardness ratio. Besides, multi-band cross-correlations were explored.
 
\section{Conclusion}
The optical studies revealed that the statistical nature of the intra-day variability can be characterized by a single power-law PSD in the Fourier space; and that the multi-band observations are strongly correlated with an occasional lead/lag of a few tens of minutes between the wave bands. The results suggests that rapid variability most likely originates in highly magnetized compact substructure of the blazar jets, and the rapid particle acceleration and cooling mechanisms can be associated with either \emph{shock-in-jet} model  or magnetic reconnection in turbulent jets.  From the distribution of the emission sizes derived from the minimum timescales, we found that some of the emission regions were smaller than the gravitational radius ($r_{g}$) of an AGN with a typical black hole mass of $\sim10^{9}\ \rm M_{\odot}$.  This can be possible either if the flux modulations occur at a fraction of the entire black hole region or in the scenario where the fluctuations reflect small-scale  magnetohydrodynamic instabilities with the turbulent structures moving relativistically in random directions. Alternatively, high  bulk Lorentz factors (e.g., $\Gamma$$\sim 100$) associated with the emitting regions, such as in \emph{ jets-in-a-jet} model,  can make the size of these regions appear comparable to $r_{g}$.  In BL Lacs, hard X-ray emission might represent the synchrotron emission from the high energy tail of the power-law distribution of the electrons. In such a case, the Lorentz factors for the highest energy electrons can be constrained as  $\sim 10^6$,  and  the variability timescales can be directly linked to the particle acceleration and cooling timescales. Whereas in FSRQs, the emission could be result of the inverse-Compton of the circum nuclear photon field (e.g. from dusty torus and broad-line region) by the low energy end of the distribution. In such a scenario, the energy of the particle turns out to be a few tens of Lorentz factors. \\

I acknowledge the financial support by the Polish National Science Centre through the grants UMO-2017/26/D/ST9/01178.


\begin{thebibliography}{}

\bibitem[Bhatta et al.(2018)]{Bhatta2018b} Bhatta, G.,  Mohorian, M., \&, Bilinsky, I.,\ 2018, \textit{A\&A} in press, (arXiv:1710.09910v2)

\bibitem[Bhatta \& Webb(2018)]{Bhatta2018a} Bhatta, G., \& Webb, J.\ 2018, \textit{Galaxies}, 6, 2

\bibitem[Bhatta et al.(2016b)]{bhatta16b} Bhatta, G., Stawarz, {\L}., Ostrowski, M., et al.\ 2016b, \textit{ApJ}, 831, 92
\bibitem[Bhatta et al.(2015)]{Bhatta2015} Bhatta, G., et al.\ 2015, \textit{ApJL}, 809, L27

\bibitem[Bhatta et al.(2013)]{bhatta13}Bhatta, G., et. al. 2013, A\&A, 558A, 92B

\end{thebibliography}
\end{document}